\documentclass{pasj00}

\begin{document}
\SetRunningHead{Shimizu and Inoue}{The Effect of the Remnant Mass in
Estimating Stellar Mass of Galaxies}

\title{The Effect of the Remnant Mass in Estimating Stellar Mass
of Galaxies}

\author{Ikkoh \textsc{Shimizu} \& Akio K. \textsc{Inoue}}
\affil{College of General Education, Osaka Sangyo University, 3-1-1
Nakagaito, Daito, Osaka 574-8530, Japan} 
\email{shimizu@las.osaka-sandai.ac.jp}

%


\maketitle

\begin{abstract}
The definition of the galactic stellar mass estimated from the spectral
 energy distribution is ambiguous in the literature; whether the stellar
 mass includes the mass of the stellar remnants, i.e. white dwarfs,
 neutron stars, and black holes, is not well described.  
The remnant mass fraction in the total (living+remnant) stellar mass of
 a simple stellar population monotonically increases with the age of the
 population, and the initial mass function and metallicity affect the
 increasing rate.  
Since galaxies are composed of a number of stellar populations, the
 remnant mass fraction may depend on the total stellar mass of galaxies
 in a complex way.
As a result, the shape of the stellar mass function of galaxies may
 change, depending on the definition of the stellar mass.
In order to explore this issue, we ran a cosmological hydrodynamical
 simulation, and then, we have found that the remnant mass fraction
 indeed correlates with the total stellar mass of galaxies. 
However, this correlation is weak and the remnant fraction can be
 regarded as a constant which depends only on the redshift.  
Therefore, the shape of the stellar mass function is almost unchanged,
 but it simply shifts horizontally if the remnant mass includes or not.
The shift is larger at lower redshift and it reaches 0.2-dex at $z=0$
 for a Chabrier IMF.  
Since this causes a {\it systematic} difference, we should take care of
 the definition of the `stellar' mass, when comparing one's result with
 others.
\end{abstract}

\section{Introduction}
The estimation of the total mass (dark matter, stellar components and
gas) in galaxies is very important to know how the galaxies evolve
through the cosmic time. The simplest way to estimate the mass is to
observe the kinematics of bright stars, star clusters and/or satellite
galaxies (e.g., \cite{Carollo2010, Beers2012, Kafle2012}). However, this
method is limited to nearby galaxies because the current telescopes do
not have enough resolution and sensitivity for high-$z$ galaxies. The
estimation of the dynamical mass of high-$z$ galaxies has been performed
by using gas motions. In addition, the halo mass of high-$z$ galaxies is
sometimes investigated by indirect method like the auto correlation
function (e.g., \cite{Gawiser2007}). On the other hand, it is difficult
to directly estimate the mass of stellar components and gas with this
method, because the dark matter is the dominant component 
and the baryonic mass contribution in galaxies is small. Thus, we need
to consider another mass estimator for the baryonic components. We can
estimate the baryonic mass using its emission in the local
Universe to high-$z$. However, there are some uncertainties in
the process of the conversion from the observational luminosity to
the mass.

Nowadays, a number of studies have been performed to investigate
the stellar mass in galaxies at various redshifts (e.g., \cite{Bell2001,
Conroy2009, Longetti2009, Wilkins2013}).
These studies are based on the population synthesis models 
such as {\scriptsize P\'{E}GASE} \citep{PEGASE}, {\scriptsize GALAXEV}
\citep{GALAXEV} and {\scriptsize Maraston} \citep{Maraston} and a
spectral energy distribution (SED) fitting method (e.g., \cite{saw98,
Schaerer2009}). 
This is now routinely adopted for the stellar mass estimation from the
nearby Universe to high-$z$. These estimators based on the population
synthesis models need the broad-band photometry from the 
ultraviolet (UV) to the near-infrared (NIR) in order to increase the
precision. Especially, the information of the rest-frame optical to NIR
band is necessary to get the reliable stellar mass. In the population
synthesis models, we need to assume an initial mass function (IMF), the
metallicity, the star formation history, the nebular emission, the dust
attenuation and the treatment of the thermally pulsing asymptotic giant
branch (TP-AGB) phase. The result based on the population synthesis
models strongly depends on these assumptions which also causes large
systemic uncertainties (e.g., \cite{Conroy2009}). 
For example, \citet{Schaerer2009} claimed that if the nebular emission
is considered in the SED fitting method, the stellar mass is a
factor of two smaller than the mass without the nebular emission.  

The definition of the stellar mass can be another source of uncertainty. 
\citet{Renzini2006} and \citet{Longetti2009} gave explicit definitions of the `stellar' mass: 
(1) the mass of living stars which are still shining (or in the nuclear burning phase); 
(2) the mass of living stars and stellar remnants 
such as white dwarfs (WDs), neutron stars (NSs), and black holes (BHs); 
(3) the mass of living stars, stellar remnants and the gas lost by stellar winds and supernovae. 
\citet{Longetti2009} actually claimed that the differences of the stellar mass among these definitions become up to a factor of two. 
Nevertheless, it does not seem that many papers about the stellar mass estimation describe which definition they used. 

The population synthesis models calculate not only the SED evolution of a galaxy 
but also the mass evolution of the stellar component composed of living stars 
and dead remnant stars of WDs,  NSs and BHs. 
In fact, the stellar remnants in galaxies do not contribute to their SEDs very much, 
but the remnants are very important for the stellar dynamics in the globular clusters \citep{GALDY}. 
The treatment of the stellar remnants may be important also for the stellar mass estimation 
if their contribution to the total (living+dead) stellar mass is not negligible. 
Indeed, the omission (or inclusion) of the remnant mass obviously causes 
a {\it systematic} effect on the stellar mass estimation.

The same problem also arises in the simulations of galaxy formation and evolution.
In the simulations, a unit of gas satisfying a set of criteria for star formation 
is converted into one (or more) star cluster(s), because we can not resolve each star. 
The mass of the star cluster(s) decreases by stellar mass loss as time passes, 
and the lost mass is returned into the gas and recycled for star formation of the next generation. 
In the cluster(s), the stars at the end of their life-time die and become remnants: 
either of WDs, NSs, or BHs, depending on the initial mass. 
The mass of the cluster(s) in the simulations should be defined as that including the remnant mass 
because both living and dead stars feel the gravity which determines the motion of the cluster(s).
However, there are two choices for the definition of the `stellar' mass for output; 
the remnant mass is included or not. 
Unfortunately, it does not seem to describe the definition explicitly in the literature. 
It is needless to say that we must take the same definition of the `stellar' mass 
when comparing simulations with observations. 

The fraction of the remnant mass in a star cluster monotonically increases with time. 
In addition, the formation rate of the stellar remnants depends on the initial mass function (IMF) and metallicity. 
This suggests that the mass fraction of the remnants in the total stellar mass 
of a galaxy depends on the star formation history and the age of the galaxy. 
Then, it may modulate the shape of the stellar mass function of galaxies, 
depending on the definition of the `stellar' mass including or not including the remnant mass, 
because there are various galaxies on various evolutionary stages. 
In general, the remnant fraction in aged galaxies is larger than that in young galaxies. 
Since more massive galaxies tend to be more passive and older than less massive galaxies, 
it may be expected that a larger fraction of the remnant for more massive galaxies. 
If it is true, we will observe a more rapid decline of the mass function towards more massive galaxies 
in the case of the stellar mass without the remnants than in the case of total one. 
Such a change of the shape of the stellar mass function of galaxies caused 
by the different definition of the `stellar' mass has not been discussed so far. 
Therefore, in this paper, we explore this issue with our hydrodynamical simulation 
of galaxy formation and evolution. 
Such a simulation is essential to treat realistic star formation histories of galaxies 
composed of a number of stellar populations with various ages and metallicities. 

We start to review the evolution of simple stellar populations (SSPs) in Section 2. 
In Section 3, we present our results of the mass fraction of living stars by using our cosmological hydrodynamic simulations.  
Our conclusions and discussions are given in Section 4. 
In this study, we concentrate on (1) the mass of living stars and (2)  the mass of living stars and the stellar remnants 
among the three definitions described above, 
because the definition (3) is too naive to be discussed. 

Throughout this paper, we adopt a $\Lambda$CDM cosmology 
with the matter density $\Omega_{\rm{M}} = 0.27$, 
the cosmological constant $\Omega_{\Lambda} = 0.73$, 
the Hubble constant $h = 0.7$ in the unit of $H_0 = 100 {\rm ~km ~s^{-1} ~Mpc^{-1}}$ 
and the baryon density $\Omega_{\rm B} = 0.046$. 
The matter density fluctuations are normalized by setting $\sigma_8 = 0.81$ \citep{WMAP}.

\section{The Remnant Mass in Simple Stellar Populations}
In this section, we consider a set of simple stellar populations 
(i.e., instantaneous-burst models) in order to gauge the maximum 
impact of the remnant mass on the total stellar mass, 
while real (or simulated) galaxies consist of many star clusters with an
IMF, an age and a metallicity.  
First, we see the difference in the functional shape of popular IMFs. 
Then, we show the time evolution of the stellar components.

We adopt two well known IMFs in this paper. 
The first is the Salpeter IMF \citep{Salpeter} and the other is the Chabrier IMF \citep{Chabrier}, 
which are shown in Fig.~1 as dashed and solid lines, respectively.
The both are normalized as $\int \phi(m) dm = 1$, where $\phi(m)$ is the IMF. 
We use the following formulae for the two IMFs:
\begin{equation}
\phi_{\rm SAL}(m) = 0.05 m^{-2.35} 
\end{equation}
\begin{eqnarray}
\phi_{\rm CHA}(m)=\left\{ \begin{array}{ll}
0.4 m^{-1} \exp{\left[- \frac{(\log{m} + 0.658)^2}{0.65}\right]} & (m< 1{\rm M_{\odot}}) \\
0.2 m^{-2.3} & (m > 1{\rm M_{\odot}})
\end{array} \right.
\end{eqnarray} 
where $\phi_{\rm SAL}(m)$ and $\phi_{\rm CHA}(m)$ are the Salpeter IMF
and the Chabrier IMF, respectively.  
The lower and upper mass limits are assumed to be $0.08$ and $100$
${\rm M_\odot}$, respectively. 
As found from Fig.~1, the Chabrier IMF is flatter than that of the
Salpeter IMF when $m<1~{\rm M_{\odot}}$.  
As a result, the Chabrier IMF has a more massive average mass of stars than the Salpeter IMF, 
and then, more stellar remnants are formed in the Chabrier IMF than in the Salpeter IMF. 
The choice of the lower mass limit of the IMFs is an issue. For example, 
if we adopt 0.1 M$_{\odot}$ instead of 0.08 M$_{\odot}$, the mass-to-light ratio 
is reduced about 10\% (1\%) for the Salpeter (Chabrier) case. On the other hand,
the evolution of the living star mass fraction, which we discuss more in the following,
does not affect so much: ($<5$\%).

\begin{figure}
\begin{center}
\includegraphics[width=80mm]{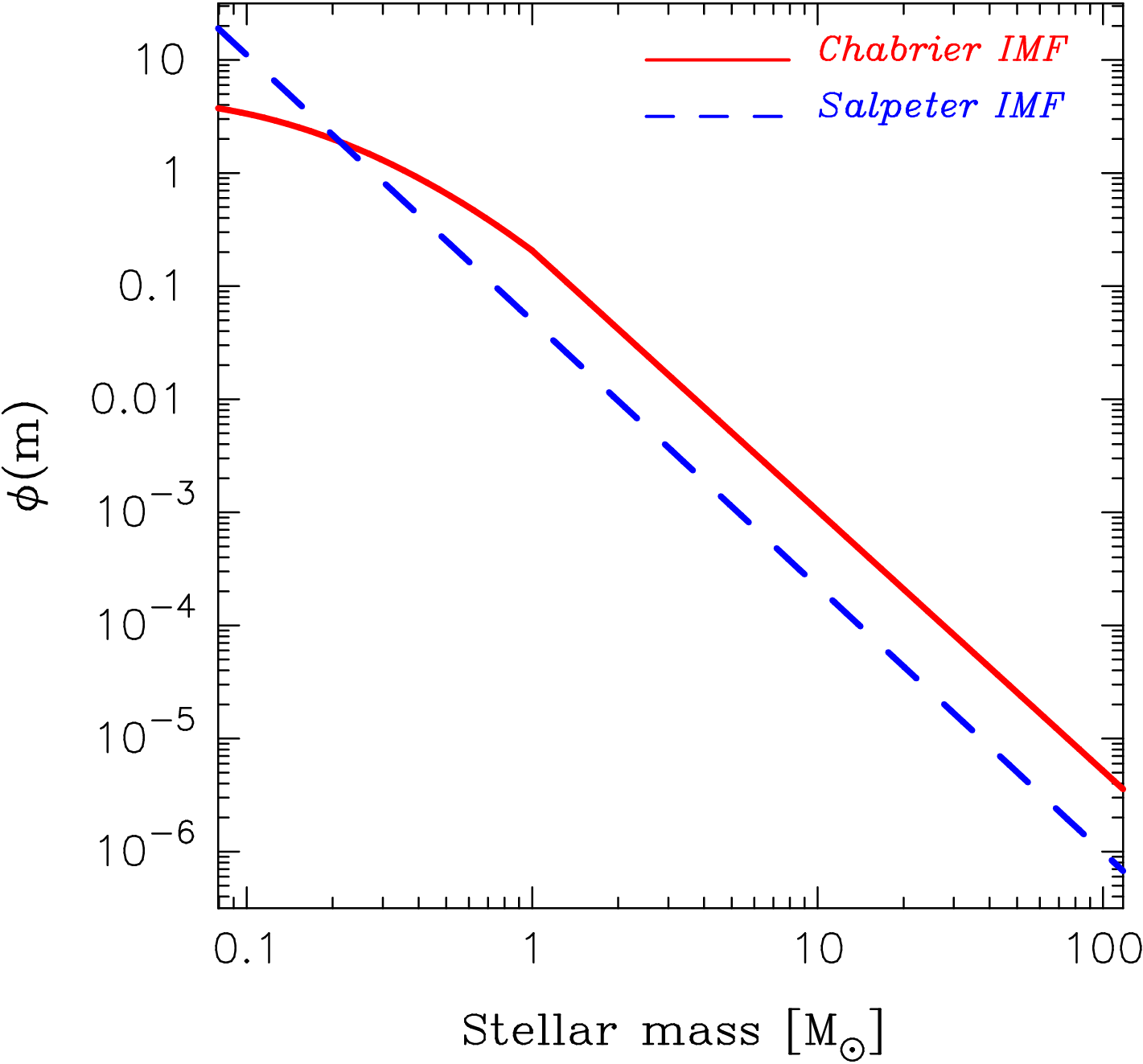}
\caption{The two initial mass functions (IMFs) assumed in this paper.
The solid and dashed lines are the Chabrier IMF \citep{Chabrier} and 
the Salpeter IMF \citep{Salpeter}, respectively. 
The both are normalized as $\int \phi(m) dm = 1$, 
where $\phi(m)$ is the IMF. }
\end{center}
\label{FIG1}
\end{figure}

Using the two IMFs, we explore the time evolutions of the mass of living stars and remnants 
(i.e., WDs, NSs and BHs) for some simple stellar populations. 
In this paper, we calculate the evolutions by using {\scriptsize P\'{E}GASE} \citep{PEGASE}. 
Fortunately, the difference in the evolution of remnant mass among the stellar population synthesis codes 
of {\scriptsize P\'{E}GASE}, {\scriptsize GALAXEV} \citep{GALAXEV} and {\scriptsize Maraston} 
\citep{Maraston} is very small ($<$ a few \%). 

\begin{figure*}
\begin{center}
\includegraphics[width=160mm]{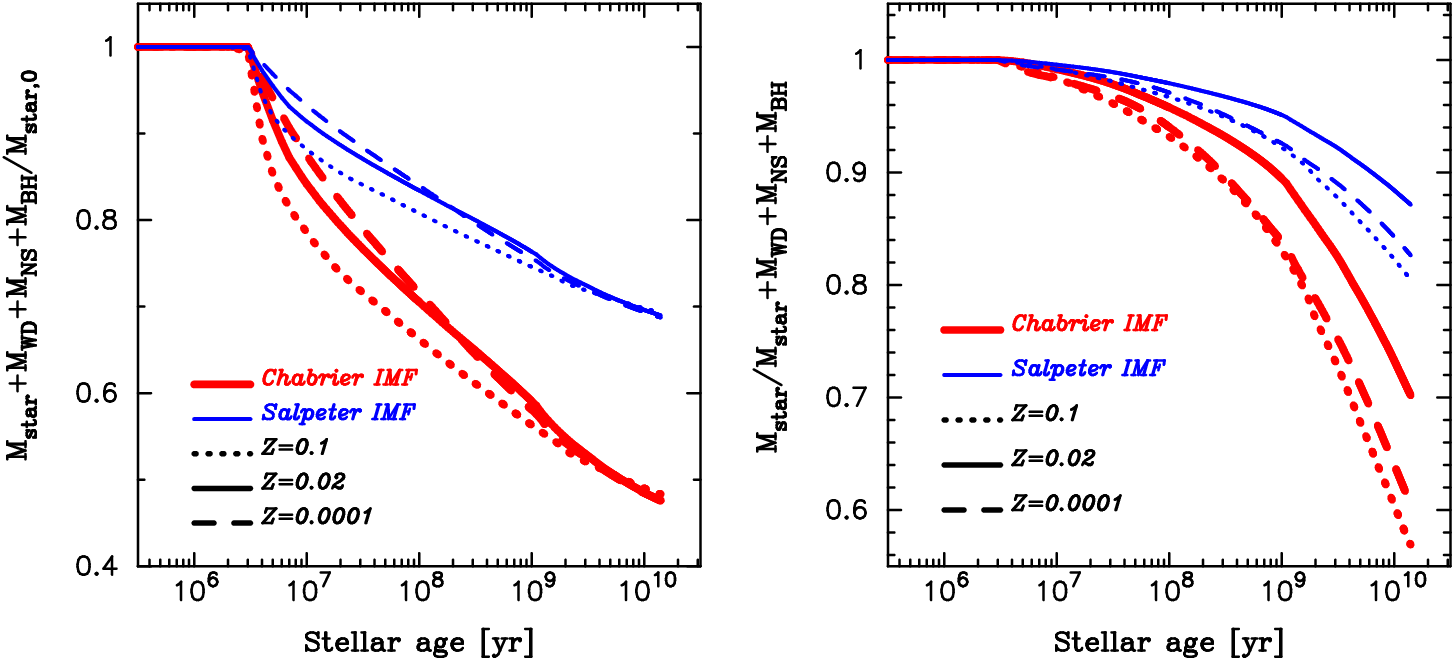}
\caption{The time evolutions of the mass of all stellar components normalized 
by the initial mass (left) and the time evolutions of the mass fraction 
of living stars in the total (living+remnant) stellar mass (right) 
for some simple stellar populations (i.e., instantaneous-burst models). 
The thick and thin lines represent the case of the Chabrier IMF and the Salpeter IMF, respectively. 
The dotted, solid and dashed lines show the case with the metallicity of 
$Z = 0.1$, $Z = 0.02~(=Z_{\odot})$ and $Z = 0.0001$, respectively. }
\end{center}
\label{FIG2}
\end{figure*}

Fig.~2 shows the time evolution of the total stellar mass normalized 
by the initial mass (left) and the time evolutions of the mass fraction of living stars 
in the total (living+remnant) stellar mass (right). 
We show the three cases of the metallicity ($Z$) as well as the two cases of the IMF. 
In the left panel, we find that the metallicity effect in the evolution
of the total (living+remnant) stellar mass is smaller than the
difference between the two IMFs. 
As expected, the total stellar mass more rapidly decreases for the Chabrier IMF case 
where an average mass of stars is larger and more short-lived. 
Note that the mass lost from the stellar components is returned into the gas phase. 

In the right panel, we find that the Chabrier IMF again predicts more rapid evolution 
of the mass fraction of the living stars (and also that of the remnants) than the Salpeter IMF. 
The stellar remnants contribute to the total (living+remnant) stellar mass 
upto 30--40\% (or 10--20\%) for the Chabrier (or Salpeter) IMF at the time of 10 Gyr. 
After that, the living (or remnant) fraction further decreases (increases) 
and will becomes zero (unity) at infinity. 
Moreover, the evolution of the living mass fraction (or the remnant mass fraction) 
also depends on metallicity; 
Among the three metallicities shown in Fig.~2, the solar
metallicity case predicts the slowest evolution of the 
living fraction in the total stellar mass of a cluster.

\begin{figure}
\begin{center}
\includegraphics[width = 80mm]{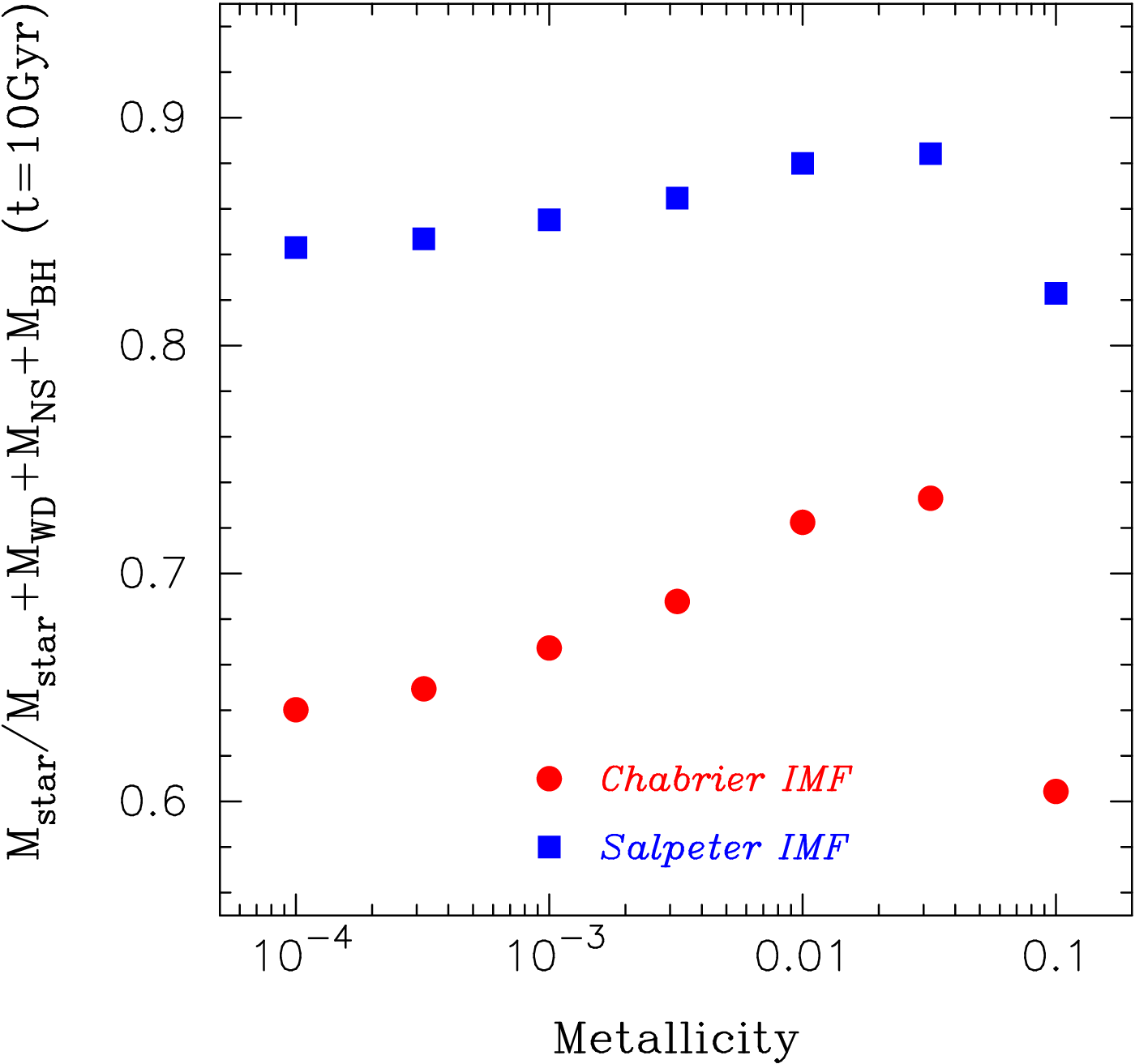}
\caption{The metallicity dependence of the mass fraction of living stars in 
the total stellar mass of a star cluster with the age of $10~{\rm Gyr}$. 
The filled circle and square points show the Chabrier IMF and the Salpeter IMF, respectively. }
\end{center}
\label{FIG3}
\end{figure}

Finally, we show the metallicity dependence of the mass fraction of living stars 
in the total (living+remnant) stellar mass more in detail. 
Fig.~3 represents the mass fraction at $10~{\rm Gyr}$ as a function of metallicity. 
We find that the fraction is proportional to the metallicity at $Z < 0.02~(= Z_{\odot})$. 
However, it suddenly drops at the metallicity more than $Z_{\odot}$ in both IMFs. 
The mass fraction of the living stars in a star cluster reaches a peak at around solar metallicity. 
At $Z < Z_{\odot}$, the mass loss of massive stars is proportional to their metallicity. 
Thus, the mass of the stellar remnant decreases with increasing the metallicity. 
On the other hand, if the metallicity of a star cluster exceeds about solar metallicity, 
the mass loss becomes efficient even for low mass stars ($\sim
1M_{\odot}$) which are still in the nuclear burning phase.   
Then, the mass fraction of living stars drops at super-solar metallicity. 
However, this is a possible answer. 
We note that the final mass of stars is not clear yet. 
This is one of the hottest topics of the stellar evolution.

\section{The Remnant Mass in Simulated Galaxies}
As seen in the previous section, the inclusion or omission of the remnant mass results 
in at most a factor of $\sim2$ {\it systematic} effect on the `stellar' mass estimation 
of galaxies with an age about 10 Gyr for the Chabrier IMF case. 
In particular, the oldest and the most metal-rich ($Z>Z_\odot$) galaxies 
may have the smallest living star fraction. Such galaxies are expected to be the most massive. 
Therefore, the shape of the stellar mass function at the massive end may be significantly different 
if we include the remnant mass in the `stellar' mass or not. 
This argument, however, may be too simple because we have examined only evolutions 
of some single star clusters whose stars all have a specific metallicity and age. 
Real galaxies are composed of a number of star clusters with different metallicities 
and ages and in some cases even different IMFs. 
Thus, we should consider the remnant impact based on a more realistic
treatment of the galaxy evolution.  
For this aim, we adopt a cosmological hydrodynamical simulation in this section. 
Hereafter, we only consider the Chabrier IMF which is more realistic 
in the local Universe than the classical Salpeter one. 

The universal IMF assumption in this paper is just simplicity
although real galaxies may have non-universal IMF (e.g., \cite{Baugh2005, vanDokkum2010}).  
Some researchers claimed that the IMF becomes top-heavy in the early
Universe and in the major-merger induced starburst (e.g., \cite{Yoshida2003, Baugh2005}). 
The remnant fraction of a top-heavy IMF is larger than those of
the Salpeter and the Chabrier IMFs.  
Moreover, the IMF may depend on galaxy types (morphology, star formation
activity, etc.) \citep{vanDokkum2010, Cappellari2012}.  
This means that the remnant fraction also depends on galaxy types. 
In these cases, the stellar mass difference among the mass definitions
becomes more prominent than in the standard IMF case which we treat
in this paper. This varying IMF issue could be an interesting future work. 

In this paper, we adopt the simulation with the same parameter set 
as our previous papers \citep{Shimizu2011, Shimizu2012}. 
However, we adopt the Chabrier IMF rather than the Salpeter IMF.
Our simulation code is based on an updated version of the Tree-PM
smoothed particle hydrodynamics (SPH) code 
GADGET-3 which is a successor of Tree-PM SPH code GADGET-2 (Springel 2005). 
We implement relevant physical processes such as star formation,
supernova (SN) feedback and chemical enrichment following
\citet{Okamoto2008, Okamoto2009, Okamoto2010}. 
Following \citet{Okamoto2010}, 
we assume that the initial wind speed of gas around stellar particles
that trigger SN is proportional to the local velocity dispersion of dark matter. 
The model is motivated by recent observations of \citet{Martin2005}. 
We also have implemented a model where gas cooling is quenched in
galaxies whose halos' velocity dispersions are greater than a threshold value to reproduce the stellar mass function at $z = 0$ (Shimizu et al. in preparation). 
We employ $N = 2 \times 256^3$ particles in a comoving volume of $40 h^{-1}{\rm ~Mpc}$ on a side. 
The mass of a dark matter particle is $2.41 \times 10^8 h^{-1}{\rm M_{\odot}}$ 
and that of a gas particle is $4.95 \times 10^7 h^{-1}{\rm M_{\odot}}$, respectively. 
This is smaller number of particles and box size than our previous work
\citep{Shimizu2011, Shimizu2012},  but the mass resolution is the same. 
We run a friends-of-friends (FoF) group finder to locate groups of stars. 
Then, we also identify substructures (subhalos) in each FoF group using SUBFIND
algorithm developed by \citet{Springel2001}. 
We regard substructures as bona-fide galaxies. 

\begin{figure*}
\begin{center}
\includegraphics[width = 160mm]{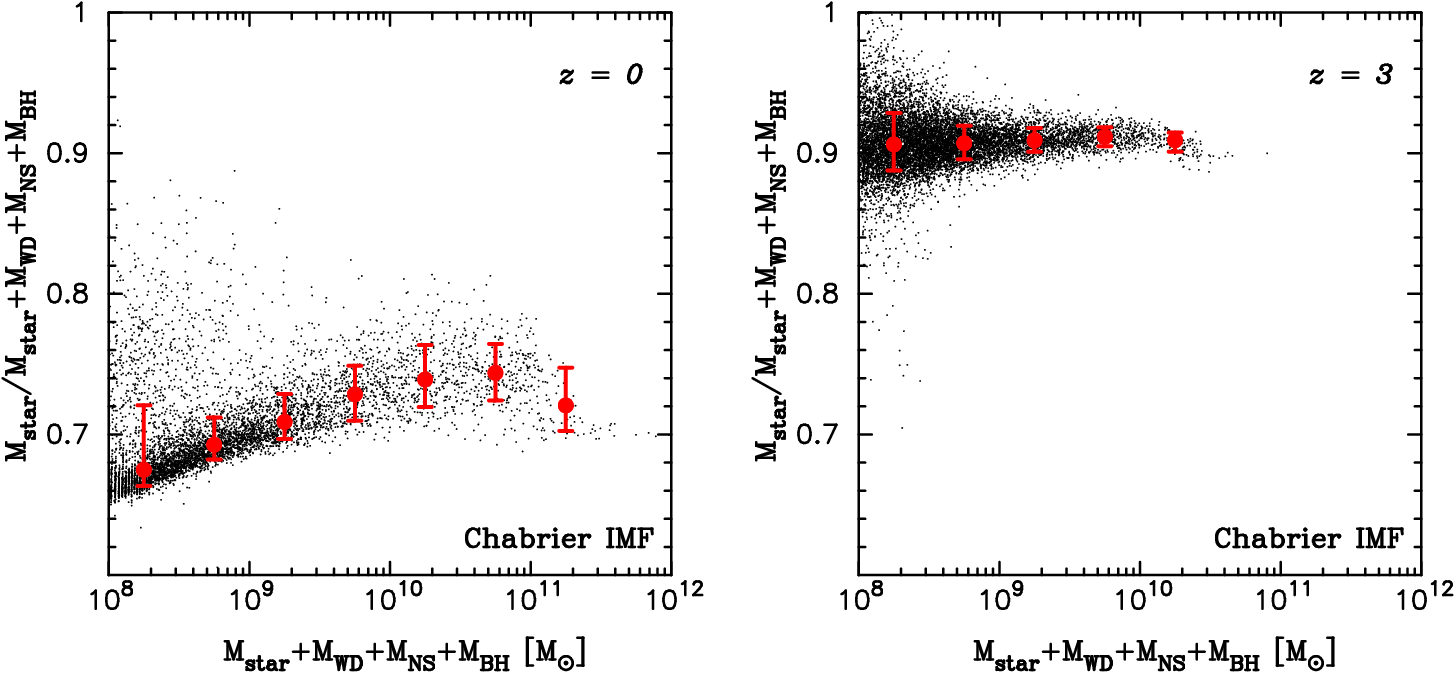}
\caption{The mass fraction of living stars in the total (living+remnant) stellar mass 
as a function of the total stellar mass of the simulated galaxies 
at $z = 0$ (left) and $z = 3$ (right). 
The small dots represent the fraction of the all simulated galaxies. 
The filled circles with error-bars represent the median and the central 68\% range in each stellar mass bin. }
\end{center}
\label{FIG4}
\end{figure*} 

\begin{figure*}
\begin{center}
\includegraphics[width = 160mm]{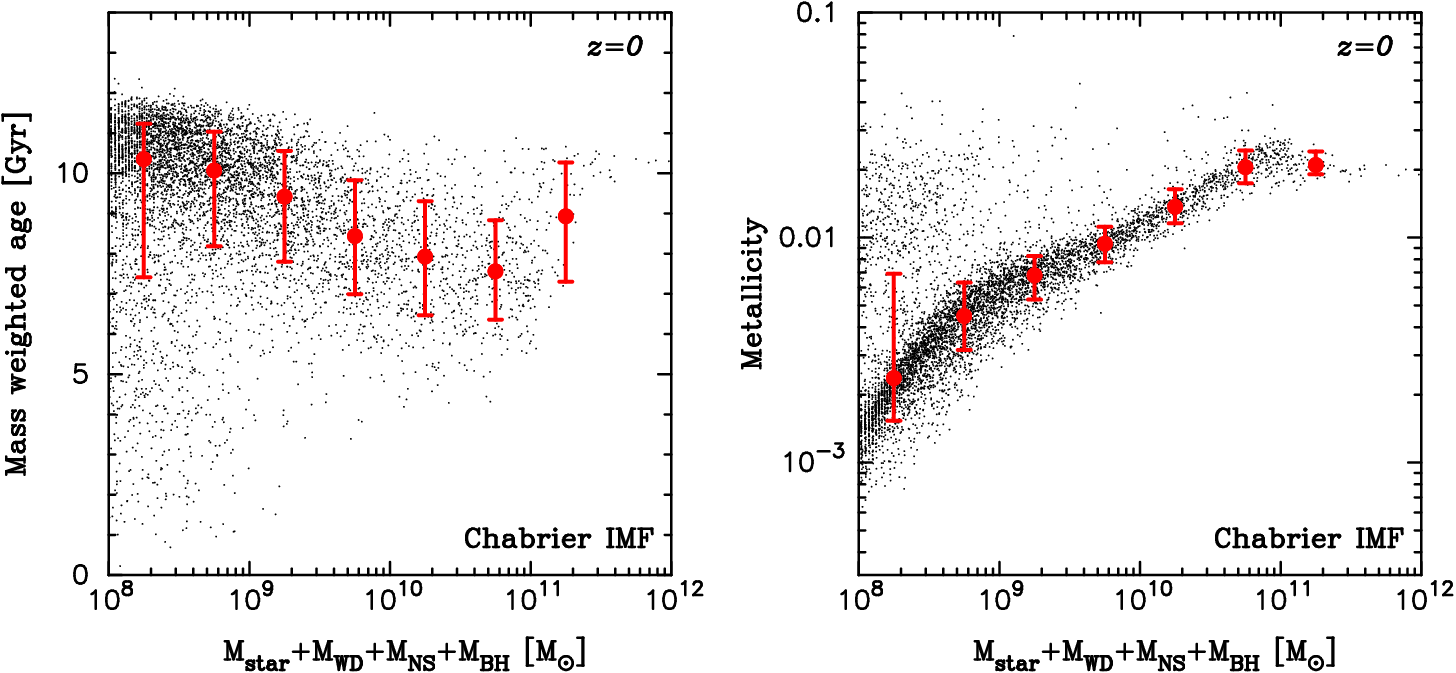}
\caption{The mass weighted average age (left) and metallicity (right) of the all stellar components 
in the simulated galaxies as a function of the total stellar mass at $z = 0$. 
The small dots show the locations of all simulated galaxies in the panels. 
The filled circles with error-bars represent the median 
and the central 68\% range in each stellar mass bin. }
\end{center}
\label{FIG5}
\end{figure*}

\begin{figure*}
\begin{center}
\includegraphics[width = 160mm]{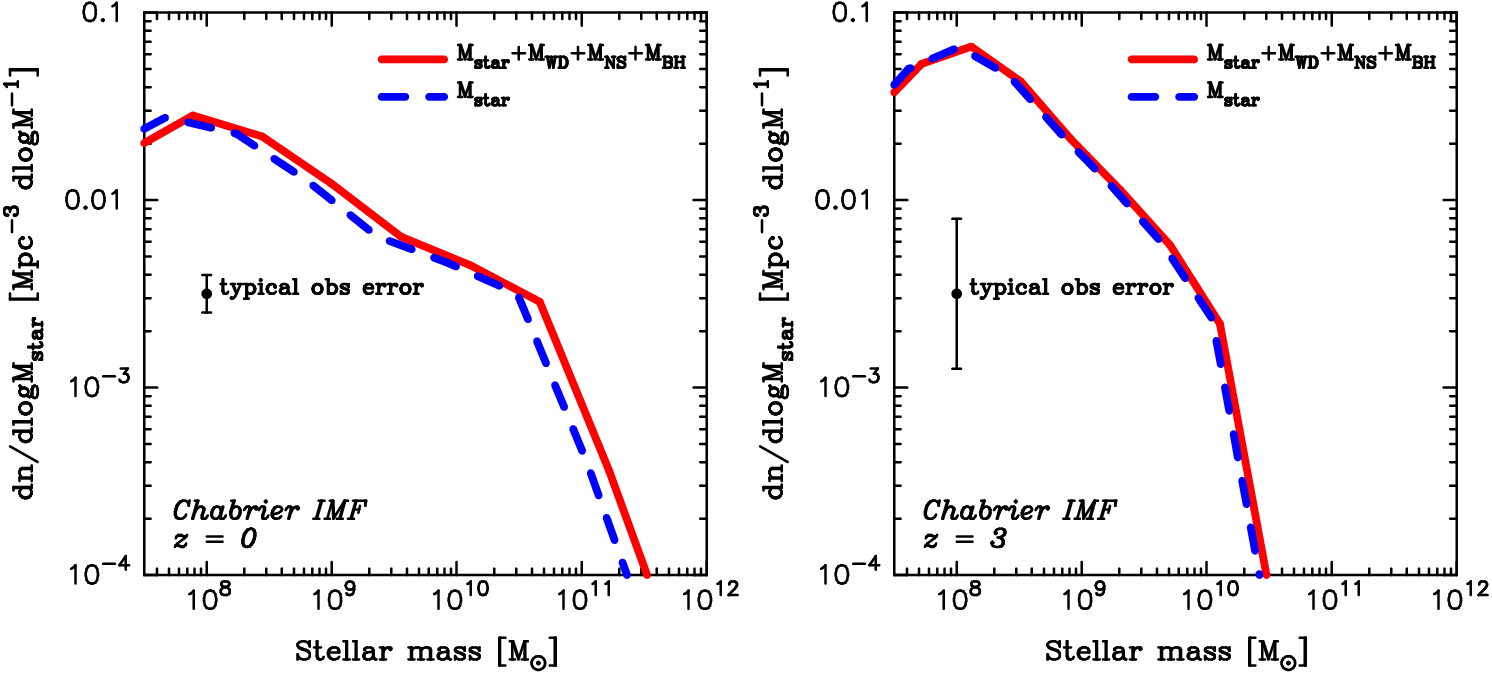}
\caption{The stellar mass function at $z = 0$ (left) and $z = 3$ (right). 
The solid and dashed lines show the functions of the sum of all stellar components 
and the case of only the living stars, respectively. 
The point with error bar is the typical observational error at each redshift. 
The error size at $z=0$ and $z = 3$ is estimated from \citet{Li2009, Baldry2012} and \citet{Elsner2008, Mortlock2011, Santini2012}. }
\end{center}
\label{FIG6}
\end{figure*}

Fig.~4 shows the mass fraction of the living stars 
relative to the total (living+remnant) stellar mass in the simulated galaxies 
at $z=0$ (left panel) and $z=3$ (right panel). 
The small dots represent the locations of the all simulated galaxies in the panels, 
while the filled circles with error-bars show the median 
and the central 68\% range in each stellar mass bin. 
We find that at $z=0$ (left panel) the living fraction slowly increases
when the total stellar mass increases, except for the massive end of $>10^{11}$ ${\rm M_\odot}$, 
where the living fraction decreases with the total stellar mass. 
Although the trend in the massive end is the same as our expectation that the fraction of the living stars 
in more massive galaxies is smaller than that in less massive galaxies, 
the trend at $<10^{11}$ ${\rm M_\odot}$ is different from and even opposite to it. 
At $z=3$ (right panel), the living fraction is almost constant. 

In order to understand the mechanism to produce these trends found in Fig.~4, 
we show in Fig.~5 the mass weighted average age (left panel) 
and metallicity (right panel) of the stellar components in the simulated galaxies at $z=0$. 
We compute the mass weighted age of each simulated galaxy as following equation, 
\begin{equation}
\tau_{\rm age} = \frac{\sum_i t_i^{\rm age} m_i^{\rm star}}{\sum_i m_i^{\rm star}}, 
\end{equation}
where $t_i^{\rm age}$ and $m_i^{\rm star}$ are the age and mass of the {\it ith} star cluster in a simulated galaxy. 
The definition of the metallicity $Z_{\rm gal}$ of simulated galaxies is 
\begin{equation}
Z_{\rm gal} = \frac{\sum_i Z_i^{\rm star} m_i^{\rm star}}{\sum_i m_i^{\rm star}}, 
\end{equation}
where $Z_i^{\rm star}$ is the metallicity of the {\it ith} star cluster in simulated galaxies.  

The situation at higher-$z$ is essentially same as that at $z=0$, 
except for the average age becomes younger at higher-$z$. 
The average age (left panel) gradually decreases with increasing the total stellar mass 
from $10^8$ M$_\odot$ to $10^{11}$ M$_\odot$. 
This is because the gas (and the dark matter) accretion into the galaxies 
and the gas cooling in the galaxies are more effective for more massive galaxies in this mass range. 
As a result, the star formation continuously occurs in such galaxies 
and the average age of stars becomes younger there. 
At the massive end ($>10^{11}~{M_{\odot}}$), on the other hand, 
the average age increases with the total stellar mass. 
This is due to quenching of the star formation activity in the mass range; 
The most massive galaxies evolve passively. 

The metallicity of the stellar components (not gas) in the simulated galaxies 
shows a clear trend; more massive galaxies are more metal-rich, 
except for the massive end ($>10^{11}~{M_{\odot}}$). 
This clear mass dependence of the metallicity at $<10^{11}$ M$_\odot$ 
is the so-called mass-metallicity relation observed well for gas metallicity \citep{Mannucci2010}. 
Note that we show here the relation of the stellar metallicity which is rarely observed so far. 
On the other hand, we find that the metallicity is almost constant 
and close to the solar value ($Z=0.02$) at the massive end $>10^{11}$ M$_\odot$. 
Interestingly, the galaxies with super-solar stellar metallicity is
very limited in our simulation. 

Now we are ready to identify the mechanism producing the trends of the living fraction in Fig.~4. 
The positive correlation between the living fraction and the total
stellar mass for $<10^{11}$ M$_\odot$ is caused by both of the age and metallicity effects. 
For $<10^{11}~{\rm M_\odot}$ galaxies, the mass wighted age of simulated
galaxies decreases with increasing the stellar mass.  
This means that the age effect is positive effect for the living fraction. 
On the other hand, the metallicity of galaxies is proportional to the stellar mass. 
This implies that the metallicity effect is positive effect for the living fraction. 
Thus, the positive correlation between the living fraction and the
stellar mass for $M<10^{11}$ M$_\odot$ is caused by both of the age and
the metallicity effects.  
However, the age range ($8 \times 10^9 ~{\rm yr} \leq \tau_{\rm age} \leq 10^{10} ~{\rm yr}$) 
and the metallicity range ($2 \times 10^{-3} \leq Z_{\rm gal} \leq 0.02$) of simulated galaxies is very small in this stellar mass range, 
so the living fraction range also small (see also Fig.~2 and Fig.~3). 
The negative correlation of the living fraction and the total stellar mass for $>10^{11}$ M$_\odot$ 
is caused only by the age effect (i.e. a passive evolution). 
In this mass range, the stellar metallicity is almost constant at the solar value, 
and then, there is no metallicity effect. 
As a result, the living fraction decreases with the stellar mass.
Note that we do not observe the rapid reduction of the living fraction 
at the super-solar metallicity found in Fig.~3 
because the stellar metallicity hardly exceeds the solar value in our simulation. 

In summary, the living stellar mass fraction in the total (living+remnant) stellar mass 
of galaxies increases with the total stellar mass for $<10^{11}$ M$_\odot$ 
and decreases for $>10^{11}$ M$_\odot$. These trends are produced 
by the star formation activity correlated with the mass 
and the mass--metallicity relation for $<10^{11}$ M$_\odot$ and 
by the passive evolution for $>10^{11}$ M$_\odot$. 
However, these trends are weak and the living mass fraction can be regarded as 
a constant independent of the total stellar mass of galaxies. 
From the comparison between the two panels in Fig.~4, 
we find that the mass fraction of living stars is larger in high-$z$ than in $z=0$. 
This is natural because the galaxies at high-$z$ are younger on average than those at $z=0$ 
and more stars are still alive at higher-$z$. 
Indeed, an average fraction of the living stars is 90\% at $z=3$; 
the remnant fraction is only 10\% at this redshift. 
On the other hand, an average living fraction at $z=0$ is about 70\%, 
so that the remnant fraction reaches about 30\%. 

Finally, we explore the effect of the stellar remnants on the shape of the stellar mass function. 
Fig.~6 shows the stellar mass function at $z=0$ (left panel) and $z=3$ (right panel). 
The solid and dashed lines represent the functions with and without the remnant mass, respectively. 
In the both panels, we also show the typical observational error at each redshift. 
Our aim of this study is that we explore whether the shape of the stellar mass function changes 
if we adopt the different definitions of the stellar mass. 
So, it is not necessarily to perfectly reproduce the observational result. 
It is enough to check whether the difference of the shape of the stellar mass function is larger than the observation error or not. 
However, if our simulation result and the observation have a large difference, 
the star formation history could be different from real one. 
Thus, a certain level of consistency between our simulation result and the observations is necessary and has been confirmed. (Shimizu et al. in prep).  

At $z=3$, the shape of the stellar mass functions with and without the
remnants are almost identical, but a slight horizontal shift can be found. 
As found in Fig.~4 right, the living fraction is 90\%, so that this shift is about 0.05-dex. 
In the $z=0$ case, the difference of the shape is also small. 
This is due to the constancy of the living fraction as a function of the
total (living+remnant) stellar mass of galaxies as shown in Fig.~4. 
However, the horizontal shift in this case is about 0.15-dex because of
the living fraction of 70\%. 
If we measure the vertical difference of the two mass functions, it is about 0.2-dex.
These horizontal and vertical shifts are indeed smaller than the current
uncertainty of the observational data (0.3-dex).
However, this is a {\it systematic} difference. 
Therefore, we should correct it when comparing one's results with others. 

\section{Summary and Discussion}

We have discussed the effect of the stellar remnants such as 
white dwarfs, neutron stars and black holes on the total stellar mass of galaxies. 
This is motivated by the fact that many papers do not  describe which definition of the stellar mass they used. 
First, we have examined the time evolutions of the mass of each stellar components for simple stellar populations
(i.e., instantaneous-burst models) with different metallicities and initial mass functions (IMFs). 
This simple treatment enables us to analyze what processes cause differences 
in the stellar mass with and without the remnants. 
Then, we have identified the IMF, metallicity, and the age 
as the sources of the differences in the remnant mass fraction. 
Namely, the remnant mass fraction monotonically increases with time. 
The Chabrier IMF, which is more realistic one in the local Universe,
predicts a factor of about 2 larger remnant mass fraction than 
the Salpeter one because of a larger average mass of stars. 
Lower metallicity also expects larger remnant mass fraction 
because of inefficient mass loss in the course of the stellar evolution. 
Real galaxies are composed of a number of stellar populations 
with various ages and metallicities (and even IMFs). 
These three effects may change the shape of the stellar mass function of galaxies, 
depending on the inclusion or omission of the remnant mass in the stellar mass. 
In order to check this issue, we have run a cosmological hydrodynamical simulation. 

We have found from our simulation results at $z=0$ that the mass fraction 
of living stars in the total (living+remnant) stellar mass of galaxies 
increases with increasing the total stellar mass 
if the total stellar mass is less than about $10^{11}$ M$_\odot$, 
and the fraction decreases if the total stellar mass exceeds about $10^{11}$ M$_\odot$. 
This positive correlation of the living fraction with the total stellar mass of $<10^{11}$ M$_\odot$ 
is produced by the following two effects: 
the negative correlation of the mass-weighted average age of the stellar component with the total stellar mass 
and the positive correlation of the mass-weighted average metallicity of the stellar components 
with the total stellar mass for $<10^{11}$ M$_\odot$. 
The former correlation means the positive correlation of the star formation activity 
with the total stellar mass. 
The latter correlation corresponds to the mass-metallicity relation for the stellar components. 
In general, the mass-metallicity relation for gas is well established at various redshifts \citep{Mannucci2010, Nagao2012, Yabe2012}. 
The relation for stellar components is rarely observed due to the observational difficulty. 
However, the trend of our simulation result is very similar to the relation for gas. 
The negative correlation of the living fraction with the total stellar mass of $>10^{11}$ M$_\odot$ 
is produced by quenching of the star formation activity in the mass range 
(i.e. a passive evolution of these galaxies). 
However, the dependence of the living mass fraction on the total stellar mass is weak. 
Therefore, we can regard the fraction as a constant independent of the total stellar mass. 
The living fraction at higher-$z$ is larger than that at $z=0$ simply because more stars are still alive at higher-$z$. 
The average living (or remnant) mass fraction is 90\% (10\%) at $z=3$ and 70\% (30\%) at $z=0$. 
Therefore, the shape of the stellar mass function does not change very much 
but just shift horizontally depending on the inclusion or omission of the remnant mass in the stellar mass 
and the amount of the shift is larger at lower-$z$. 
The shift is about 0.2-dex at $z=0$ for the Chabrier IMF. 
This causes a {\it systematic} effect and we should avoid such a difference 
when we compare one's stellar mass function with others. 
Therefore, it is important to consider the definition of the `stellar' mass. 

The IMF produces a large difference in the remnant formation rate. 
Thus, the change of the IMF along time and environment can cause a significant effect 
on the living mass fraction in the total stellar mass of galaxies. 
In particular, a top-heavy IMF, which is expected in the early Universe 
and in the starburst activity \citep{Bromm2002, Yoshida2003, Baugh2005, Prantzos2006, Maness2007, Iwata2009},
predicts much more remnant fraction in much shorter timescale. 
This suggests a possibility that the living mass fraction in the total stellar mass 
becomes significantly small even at high-$z$. 
Moreover, if the IMF changes in various galaxy types as reported by recent studies \citep{vanDokkum2010, Cappellari2012}, 
the living mass fraction also depends on the galaxy type. 
Then, it may modulate the shape of the stellar mass function, 
depending on the inclusion or omission of the remnant mass in the `stellar' mass. 
Therefore, the definition of the `stellar' mass should be described explicitly in papers 
where people present their estimations of the stellar mass.


\bigskip
We are grateful to Cheng Li, Ken Nagamine, Marcin Sawicki, 
Masaru Kajisawa, Masayuki Tanaka, Paola Santini, Pratika Dayal and 
Takashi Moriya for useful discussion about this work. 
We also thank to the anonymous referee for her/his constructive comments. 
Numerical simulations have been performed with the EUP, PRIMO 
and SGI cluster system installed 
at the Institute for the Physics and Mathematics of the Universe, 
University of Tokyo and with T2K Tsukuba at Center for Computational
Sciences in University of Tsukuba. 
We acknowledge the financial support of Grant-in-Aid for Young
Scientists (A: 23684010) by MEXT, Japan.
\bigskip


\end{document}